# Tweeting biomedicine: an analysis of tweets and citations in the biomedical literature

Stefanie Haustein[1], Isabella Peters[2], Cassidy R. Sugimoto[3], Mike Thelwall[4], & Vincent Larivière[5]

[1] *stefanie.haustein@umontreal.ca*
École de bibliothéconomie et des sciences de l'information, Université de Montréal
C.P. 6128, Succ. Centre-Ville, Montréal, QC. H3C 3J7 (Canada) and
Science-Metrix Inc., 1335 Avenue du Mont-Royal E, Montréal, Québec H2J 1Y6, (Canada)

[2] *isabella.peters@hhu.de*
Institute for Language and Information, Department of Information Science, Heinrich Heine University, Universitätsstr. 1, 40225 Düsseldorf (Germany)

[3] *sugimoto@indiana.edu*
School of Informatics and Computing, Indiana University Bloomington
1320 E. 10th St. Bloomington, IN 47401 (USA)

[4] *m.thelwall@wlv.ac.uk*
School of Technology, University of Wolverhampton, Wulfruna Street, Wolverhampton
WV1 1LY (UK)

[5] *vincent.lariviere@umontreal.ca*
École de bibliothéconomie et des sciences de l'information, Université de Montréal
C.P. 6128, Succ. Centre-Ville, Montréal, QC. H3C 3J7 (Canada) and
Observatoire des sciences et des technologies (OST), Centre interuniversitaire de recherche sur la science et la technologie (CIRST), Université du Québec à Montréal
CP 8888, Succ. Centre-Ville, Montréal, QC. H3C 3P8, (Canada)

## Abstract

Data collected by social media platforms have recently been introduced as a new source for indicators to help measure the impact of scholarly research in ways that are complementary to traditional citation-based indicators. Data generated from social media activities related to scholarly content can be used to reflect broad types of impact. This paper aims to provide systematic evidence regarding how often Twitter is used to diffuse journal articles in the biomedical and life sciences. The analysis is based on a set of 1.4 million documents covered by both PubMed and Web of Science (WoS) and published between 2010 and 2012. The number of tweets containing links to these documents was analyzed to evaluate the degree to which certain journals, disciplines, and specialties were represented on Twitter. It is shown that, with less than 10% of PubMed articles mentioned on Twitter, its uptake is low in general. The relationship between tweets and WoS citations was examined for each document at the level of journals and specialties. The results show that tweeting behavior varies between journals and specialties and correlations between tweets and citations are low, implying that impact metrics based on tweets are different from those based on citations. A framework utilizing the coverage of articles and the correlation between Twitter mentions and citations is proposed to facilitate the evaluation of novel social-media based metrics and to shed light on the question in how far the number of tweets is a valid metric to measure research impact.

## Introduction

Data from social media platforms have recently been exploited to measure early impact or types of research impact for scholarly publications in ways that complement traditional citation-based indicators. So-called "altmetrics" (Priem, 2010; Priem, Costello, & Dzuba, 2011) reflect, primarily, activity in social media environments with the purpose of gathering previously invisible traces of scholarly impact. Activities on platforms such as Mendeley, CiteULike, ResearchGate, Academia.edu, LinkedIn, Facebook, and Twitter can be tracked to rapidly monitor the manner in which scholarly documents are disseminated and discussed (Li, Thelwall, & Guistini, 2012; Piwowar, 2013; Priem & Costello, 2010). Studies of the role of social media in scholarly communication have investigated their use in diffusion (Darling, Shiffman, Côté, & Drew, 2013), conference chatter (Weller, Dröge, & Puschmann, 2011), crowdsourcing (Ogden, 2013), science popularization (Sugimoto & Thelwall, 2013), and promotion of scholarly products (Cronin, 2013; Nature Chemistry, 2013). New tools to facilitate the use of altmetrics have been introduced (e.g., Kaur, Hoang, Sun et al., 2012), research councils are encouraging the use of altmetrics for evaluative purposes (e.g., Viney, 2013), and scholars are arguing for inclusion of altmetrics on curricula vita (Piwowar & Priem, 2013). However, large-scale studies of altmetrics are rare, and systematic evidence about the reliability, validity, and context of these metrics is lacking (Wouters & Costas, 2012; Liu & Aidle, 2013). Furthermore, we lack evidence about the actors and stakeholders in both the creation and consumption of these metrics.

To further this conversation, we examine the extent to which biomedical papers are represented on Twitter and the relationships between tweets and citations for these papers. We select Twitter as one of the most popular social media websites, claiming over 200 million active users in March of 2013 (Wickre, 2013). While scientometric analyses have typically focused on measuring scholarly communication in a closed community of researchers who read, cite, and publish, altmetrics claim to capture impact measures from a broader audience. Studies are needed to systematically examine the integration of social media sources into scholarly communication, how far researchers (or other stakeholders such as science journalists, public outreach officers, journal publishers) use them to communicate research results and which audiences they target (e.g., the scientific community or an interested public). Given that tools such as Altmetric.com and ImpactStory.org[i] provide easy access to altmetrics, a systematic study of the meaning and validity of altmetrics as indicators of scholarly and/or public impact is both timely and appropriate. This paper contributes to the assessment of altmetrics by evaluating the use and discussion of biomedical publications on Twitter from a quantitative point of view.

Only a handful of studies have examined Twitter use among scholars (Priem & Costello, 2010; Weller, Dröge, & Puschmann, 2011), but most of these have concluded that Twitter is not considered a particularly important tool for scholarly dissemination. Although Thelwall, Haustein, Larivière, & Sugimoto (2013) found that Twitter data were more extensive than that from other social media sources, social media usage does not contribute to a scholar's reputation (Cruz & Jamias, 2013). Less than 10% of researchers take advantage of microblogging (Rowlands, Nicholas, Russell, Canty, & Watkinson, 2011), and a mere 2.5% of scientists are active on Twitter (Priem et al., 2011).[ii] This latter figure contrasts sharply with the 2011 estimate by eMarketer (2011) that 8.7% of the adult US population was active on Twitter. This difference might be due to the divergence of various user groups on Twitter—for example, it is common to use Twitter for complaining about newly bought products (Jansen, Zhang, Sobel, & Chowdury, 2009). In addition, the different age distribution of scholars and the general population may be a factor. However, when scholars tweet, nearly 50% of their tweets are related to scholarly communication (Holmberg & Thelwall, 2013; Chretien et al., 2011).

When tweeting, users apply several communicative devices including hashtags, directed @messages, and retweets (boyd, Golder, & Lotan, 2010). Weller and Peters (2012) suggested that retweets should be considered to be "internal citations," whereas "external citations" appear when Twitter users link to outside information (such as links to particular websites or documents). Various other terms have been proposed to describe scholars' use of Twitter in referencing scholarly publications, including "tweetations" (Eysenbach, 2011) and "citation tweets" (Priem & Costello, 2010). Twitter can be a powerful tool for sharing pointers (i.e. links) to information (boyd et al., 2010; Suh, Hong, Pirolli, & Chi, 2010). Interestingly, tweets are more likely to be retweeted when they contain links. This aspect has already been recognized by scholars and is now used for popularizing tweets with scientific content: Nearly a third of scientists' tweets contain URLs (Peters, Beutelspacher, Maghferat, & Terliesner, 2012), compared to only 22% for the general population of tweets (boyd et al., 2010). According to Priem and Costello (2010), 6% of 2,322 tweets with URLs published by scientists forwarded users to scholarly publications, either directly or via different channels such as websites, while in Holmberg and Thelwall's (2013) sample of scientists' tweets it was 2.2%. The highest proportion of tweets containing URLs (55%) was found by Weller and Puschmann (2011), whose study population consisted of approximately 600 academic users. The proportion of scientists' tweets containing links can vary between disciplines (62% to 75%; Holmberg & Thelwall, 2013). The most common link destinations are blogs, advanced Twitter services (e.g., Twitpic[iii]), or other media outlets, such as newspapers or video sharing platforms (Holmberg & Thelwall, 2013; Peters et al., 2012; Weller & Puschmann, 2011; Weller et al., 2011). The degree to which traditional scholarly practices are reflected in Twitter use (e.g., citing scientific papers in tweets or referencing via links) is key to understanding the involvement that scholars have with Twitter as a dissemination tool. A large-scale study that systematically evaluates the degree to which scholarly articles are distributed on Twitter is lacking.

The aim of this study is to analyze the extent to which biomedical publications are mentioned on Twitter by evaluating the share of tweeted documents and the average number of tweets per article by discipline, specialty, and journal. Twitter provides an opportunity to evaluate a rapid dissemination vehicle differentiating it from citations, which take much longer to accumulate. We selected biomedicine as the domain for two reasons. The first is the early adoption of social media tools generally, and Twitter, in particular, into the work of practicing physicians (Berger, 2009; Cohen, 2009; Parker-Pope, 2009). This adoption has resulted in modifications to the American Medical Association's code of ethics to include policies on social media use. As argued by Chretien, Azar, and Kind (2011, p. 566), "the existence of social media is transforming way physicians communicate with the public"; however, it is not clear whether this transformation applies to biomedical researchers as well as practitioners. A demonstrated incentive for this exists, as some studies have shown that papers with complementary knowledge diffusion obtain higher citations rates (Gargouri, et al., 2010; Henneken & Accomazzi, 2011; Lippi & Favaloro, 2012; Mounce, 2013).

The results of our research can be used as a starting point for additional analyses in order to determine how, why, by whom, and to whom scholarly documents are tweeted. The present study analyzes those tweets that mention at least one of the 1.4 million scholarly documents indexed in PubMed. The following research questions, divided into three areas, are investigated:

1) Twitter coverage of biomedical papers (PubMed):
    a. What proportion of articles, journals, specialties and disciplines indexed in PubMed are mentioned on Twitter?
    b. Which PubMed journals, specialties and disciplines are tweeted most frequently?
2) Twitter impact of biomedical papers:
    a. What is the average number of tweets per article, journal, specialty and discipline?
    b. What is the relationship between the coverage of Twitter and the number of tweets per article?
3) Comparison of tweets and citation metrics for biomedical papers:
    a. Does citation behavior on Twitter resemble citation behavior of scholars? Do tweet counts and citations correlate?

## Methods

***Data***. The dataset analyzed comprises all articles and reviews indexed in PubMed as well as in the Web of Science (WoS). This combined dataset represents the core of the biomedical literature for the 2010 to 2012 period. The link between PubMed and WoS allowed for the calculation of the number of citations received by each article. This resulted in a set of 1,431,576 documents. Citations covered those received until the end of 2012 providing different lengths of citation windows depending on the publication year of the article: that is, articles published in 2010 had more time to accumulate citations. Tweet counts for PubMed articles were are based on a previous study by Thelwall et al. (2013) and obtained from Altmetric.com, which monitors social media mentions from various sources. The tweets were retrieved by Altmetric.com between July 2011 and December 2012 and were limited to those tweets published during that time that contained a unique identifier (e.g., a PMID, a Digital Object Identifier [DOI], or a URL associated with a scholarly publisher's website) referring to the PubMed documents published during the 2010 to 2012 period. Search results are verified by cross-checking metadata from links in tweets with bibliographic information of scholarly documents.

***Analysis***. The first set of analyses focused on the degree to which articles found in both PubMed and WoS were tweeted, examining the degree to which articles were cited including variations over time. For this analysis, all 1.4 million documents were used. Twitter citation rates, that is, the mean number of tweets per article, were calculated and the distribution of tweets over articles is presented. The 15 most frequently tweeted articles are listed.

The second set of analyses focused on the 5,251 journals which were represented in the set of 1.4 million documents. Of these, 4,215 were tweeted at least once (80%). However, the PubMed coverage was extremely low for some of the journals, as only some of their papers were relevant to the biomedical field, (e.g., *Cartographic Journal*, *Language & Communication*, *Italian Studies*, *Scottish Journal of Political Economy*). Therefore, the sample for the journal analysis was limited to those journals which had 1) at least 30 papers indexed in PubMed and: 2) either a) 100 articles or reviews indexed in Web of Science between 2010 and 2012, or b) at least 70% of the total articles for that journal covered in PubMed. This selection thus excludes journals with too few papers for reliable statistics and those with low coverage in PubMed. The exclusion process left 3,812 journals. The percentage of tweeted documents (here called "Twitter coverage", $P\%_{tweeted}$) and the mean number of tweets per tweeted article (called "Twitter citation rate", $T/P_{tweeted}$) was calculated. Spearman correlations were calculated between Twitter citation rates of journals and traditional bibliometric journal indicators (i.e., Impact Factor, Eigenfactor, Article Influence score, and Immediacy Index) on the journal level.

The last set of analyses evaluated Twitter use on the level of disciplines and specialties. The 1.4 million documents were classified using the NSF journal classification system. Results are shown for only those specialties where PubMed coverage exceeded 50% in order to guarantee representativeness of the particular specialty. Similar to the journal level analysis, Twitter citation rates and Twitter coverage were calculated on the level of disciplines and specialties. Additionally, the relationship between tweets and citations was examined. The most common means of identifying the relationship between two metrics is to determine the statistical correlation between them. This has been a common approach in scientometrics, in validating new metrics, by examining the degree to which they relate with previously accepted metrics. However, as suggested by Thelwall and colleagues (2013), correlation coefficients might not be suitable for comparing altmetric and citation indicators for documents published in different time periods, as the analysis can be biased by citation delays and changes in social media use. The correlations in Table 1 confirm these biases, as Spearman's ρ is highest for articles published in 2011, where both biases are smallest. Therefore, only articles published in 2011 were used for the correlation analysis.

The top 25 journals according to Twitter citation rate, i.e., the most tweets per article on average, are listed together with information on official Twitter accounts, i.e. number of tweets and followers. This information was manually collected by Google and Twitter searches in April 2013. Number of followers and tweets were also collected for the 13 journals with Twitter coverage above 50%.

**Table 1.** Statistics and Spearman correlation of number of tweets (T) and citations (C) per document and year of publication (2010 to 2012) for papers which were mentioned on Twitter at least once (134,929 of the 1,431,576 PubMed articles).

| | N | Spearman's ρ | Mean | Median | Max. |
|---|---|---|---|---|---|
| $T^{2010}$ | 13,763 | .104** | 2.1 | 1 | 237 |
| $C^{2010}$ | | | 18.3 | 7 | 3,922 |
| $T^{2011}$ | 63,801 | .183** | 2.8 | 1 | 963 |
| $C^{2011}$ | | | 5.7 | 2 | 2,300 |
| $T^{2012}$ | 57,365 | .110** | 2.3 | 1 | 477 |
| $C^{2012}$ | | | 1.3 | 0 | 234 |
| $T^{2010\text{-}2012}$ | 134,929 | .114** | 2.5 | 1 | 963 |
| $C^{2010\text{-}2012}$ | | | 5.1 | 1 | 3,922 |

** Correlation is significant at the 0.01 level (2-tailed).

***Limitations***. Replication is a key principle of scientific research. Altmetric research faces many hurdles in this regard. First, data providers may change or become obsolete quickly, making replications impossible (Liu & Aidle, 2013). Moreover, it is still difficult if not impossible to collect complete data if no direct link (i.e., URL, PMID or DOI) between research results and their mention in other publications (e.g., videos or newspaper articles) is given; in addition, many versions of one research paper are available on the Web (see also the notion on identity resolution in Buschman and Michalek [2013]). Accordingly, research on altmetrics must find ways to combine all available versions of a document in order to form reliable indicators. We relied on the tweet counts computed by Altmetric.com which are obtained through searches based on multiple document identifiers (see Methods section). Beside the technical problems associated with collecting altmetrics, there is another crucial factor associated with their use in the evaluation of authors, papers, journals, or disciplines: the critical mass of both available documents and users contributing to the data.

A general problem of social media-based analyses is that of data reliability. Although most social media services provide APIs to make usage data accessible, we still do not know if it is possible to collect every tweet, if there are missing data, or what effects download or time restrictions have on available data. In addition, the Altmetric coverage of Twitter may be incomplete due to technical issues, such as server or network downtime. Moreover, an article may be tweeted in a way that is not easily automatically identified (e.g., "See Jeevan's great paper in the current Nature!"). Due to these limitations, we assume that our findings on coverage and usage of medical documents on Twitter are rather conservative. However, we also believe that the aforementioned limitations are likely to affect all journals, specialties, and disciplines in broadly the same way, even though individual journal issues might be disproportionately affected if they were published at the time of an a service outage with the result that an initial surge of tweets at the time of publication may have been missed.

## Results

***Articles.*** 9.4% (134,929 of the 1,431,576 documents) of the PubMed/WoS articles were tweeted at least once. There was significant variation by time: 2.4% of the papers published in 2010 were tweeted at least once, 10.9% in 2011; and 20.4% of the articles published in 2012 received at least one tweet. There were 340,751 tweets mentioning 134,929 unique articles, providing a global Twitter citation rate of 2.5 (0.2 including untweeted documents). The distribution of tweets per document is positively skewed, with 63.0% of documents only mentioned once. The most frequently tweeted document was mentioned 963 times (Table 2). The 15 most frequently tweeted papers are shown in Table 2. These mostly appeared in general science journals such as *Nature*, *Science*, *PNAS* or prestigious medical journals such as the *New England Journal of Medicine* and *Lancet*. Articles from some specialized journals were also frequently tweeted. An attempt to classify these articles as regards their topics and potential reasons for being so popular on Twitter shows that the documents are curious or funny (e.g., rank #7; #11 in Table 2), has potential health applications (#5; #6) or refers to a catastrophe such as the two most frequently tweeted articles: the article by Hess and colleagues (#1) describes the genetic alterations caused by the Chernobyl accident and was published in May 2011 shortly after Fukushima, making the subject even more topical, and the paper by Yasunari and colleagues (#2) analyzes the soil contamination in Japan caused by Fukushima. The articles by Newman and Feldman (#8) and Mottron (#12) are examples of topics that concern the whole scholarly community, i.e. Open Access and a group of researchers with autism.

**Table 2.** The 15 most frequently tweeted articles.

| Bibliographic information | Journal | Number of tweets | Rank |
|---|---|---|---|
| Hess et al. (2011). Gain of chromosome band 7q11 in papillary thyroid carcinomas of young patients is associated with exposure to low-dose irradiation | PNAS | 963 | #1 |
| Yasunari et al. (2011). Cesium-137 deposition and contamination of Japanese soils due to the Fukushima nuclear accident | PNAS | 639 | #2 |
| Sparrow et al. (2011). Google Effects on Memory: Cognitive Consequences of Having Information at Our Fingertips | Science | 558 | #3 |
| Onuma et al. (2011). Rebirth of a Dead Belousov–Zhabotinsky Oscillator | Journal of Physical Chemistry A | 549 | #4 |
| Silverberg (2012): Whey protein precipitating moderate to severe acne flares in 5 teenaged athletes | Cutis | 477 | #5 |
| Wen et al. (2011). Minimum amount of physical activity for reduced mortality and extended life expectancy: a prospective cohort study | Lancet | 419 | #6 |
| Kramer (2011). Penile Fracture Seems More Likely During Sex Under Stressful Situations | Journal of Sexual Medicine | 392 | #7 |
| Newman & Feldman (2011). Copyright and Open Access at the Bedside | New England Journal of Medicine | 332 | #8 |
| Reaves et al. (2012). Absence of Detectable Arsenate in DNA from Arsenate-Grown GFAJ-1 Cells | Science | 323 | #9 |
| Bravo et al. (2011). Ingestion of Lactobacillus strain regulates emotional behavior and central GABA receptor expression in a mouse via the vagus nerve | PNAS | 297 | #10 |
| Park et al. (2012). Penetration of the Oral Mucosa by Parasite-Like Sperm Bags of Squid: A Case Report in a Korean Woman | Journal of Parasitology | 293 | #11 |
| Mottron (2011). Changing perceptions: The power of autism | Nature | 274 | #12 |
| Villeda et al. (2012). The ageing systemic milieu negatively regulates neurogenesis and cognitive function | Nature | 271 | #13 |
| Merchant et al. (2011). Integrating Social Media into Emergency-Preparedness Efforts | New England Journal of Medicine | 267 | #14 |
| Ho et al. (2011). A Low Carbohydrate, High Protein Diet Slows Tumor Growth and Prevents Cancer Initiation | Cancer Research | 261 | #15 |

***Journals***. Of the 3,812 journals meeting the thresholds (see Methods section) 3,725 (97.7%) had at least one tweeted publication. The mean number of tweets per journal is 88.7. The most frequently tweeted journal is *Nature*, with 13,430 tweets linking to its articles (Table 3). With 41.9 tweets per document, the journal with the highest mean Twitter citation rate is *Cutis*. As can be seen in Figure 1, which shows Twitter citation rate and coverage for the 3,725 journals with at least one tweet, the majority of journals (67.4%) have had less than 20% of their content tweeted and were tweeted less than twice per tweeted document.

One possible artifact could arise from journals employing someone to tweet their own material. To investigate this, we visited the journal home page for those 13 journals with Twitter coverage above 50% and found that 9 of these had an official Twitter account and the remaining four were represented by the official Twitter accounts of their respective publishers or affiliated societies.

As the data show, an official Twitter publication policy does not necessarily lead to a higher active uptake and redistribution of articles by Twitter users. The *Journal of Addiction Medicine* ($P\%_{tweeted}$=85.1%; $T/P_{tweeted}$=1.7), *Current Opinion in Endocrinology* (74.3%; 1.1), *Ultrasound Quarterly* (67.6%; 1.1), *Simulation in Healthcare* (65.6%; 1.5), *Clinical Obstetrics and Gynecology* (64.7%; 1.1), *Psychosomatic Medicine* (63.4%; 1.8), *Current Opinion in Cardiology* (61.2%; 1.1), *Journal of Nursing Administration* (58.0%; 1.3), *Cardiology Review* (55.6%; 1.2) and the *European Heart Journal-Cardiovascular Imaging* (51.7%; 1.6) had more than half of their content tweeted but overall the journals are tweeted less frequently than average. The exceptions among those with high coverage are *Palliative Medicine* (67.9%), the *British Dental Journal* (60.4%) and *Homeopathy* (52.4%), with Twitter citation rates of 4.6, 3.2 and 6.1, respectively. With a high Twitter coverage and Twitter citation rates above average, these three journals are both broadly and intensely distributed on Twitter.

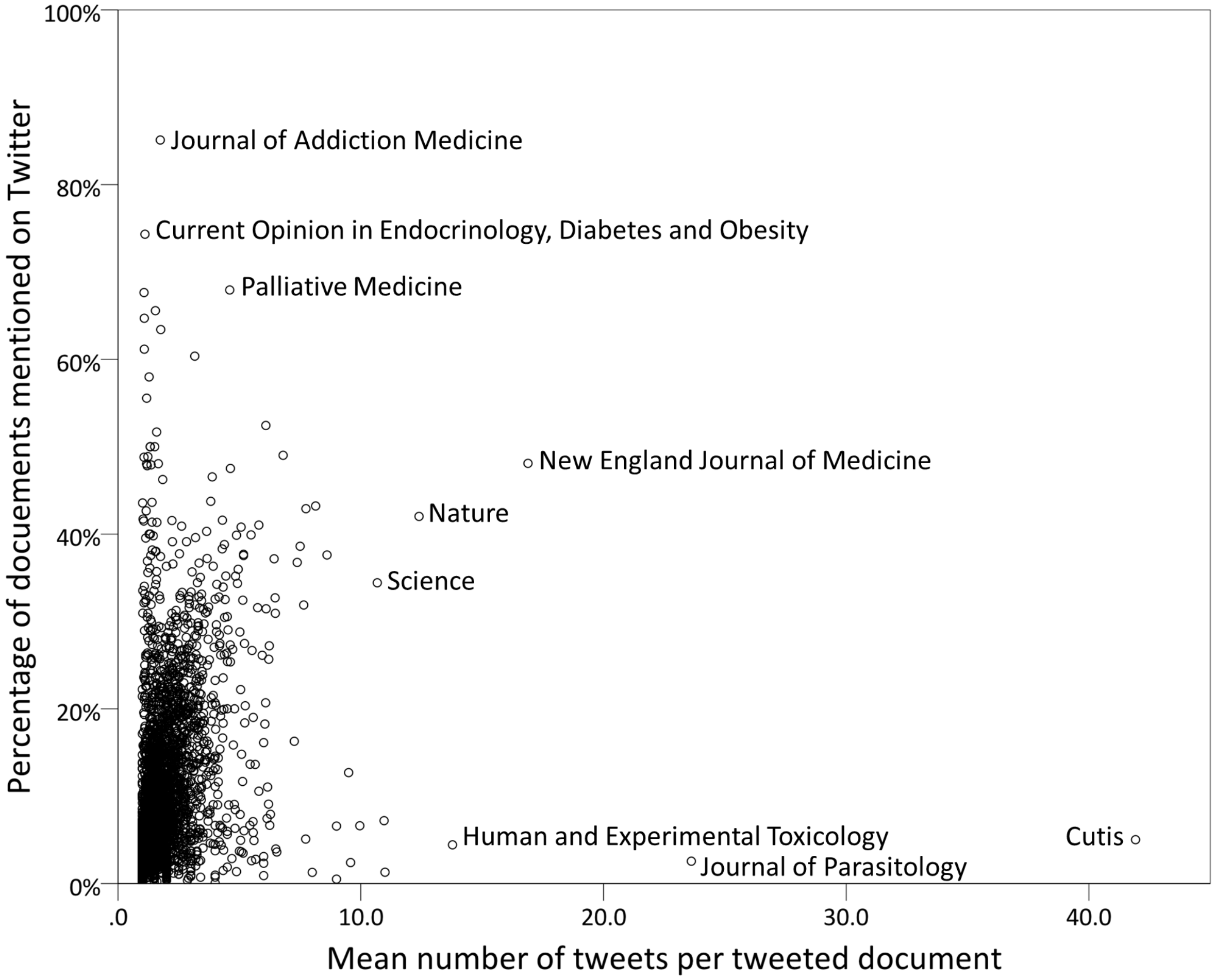

**Figure 1.** Percentage of tweeted articles (Twitter coverage) and mean number of tweets per tweeted article (Twitter citation rate) for 3,725 journals.

**Table 3.** Number of papers (articles and reviews published between 2010 and 2012) covered by PubMed and WoS (*P*), number of tweeted documents ($P_{tweeted}$), Twitter coverage ($P\%_{tweeted}$), number of tweets (*T*), mean Twitter citation rate (tweets per tweeted article, $T/P_{tweeted}$), standard deviation, median and maximum of tweets per document, journal Impact Factor 2011 ($IF^{2011}$) and information on official Twitter account (tweets/followers) for the 25 journals with the highest mean Twitter citation rate.

| Journal | P | $P_{tweeted}$ | $P\%_{tweeted}$ | T | $T/P_{tweeted}$ | Std. Dev. | Max. | $IF^{2011}$ | Twitter account (tweets / followers) |
|---|---|---|---|---|---|---|---|---|---|
| Cutis | 239 | 12 | 5.0% | 503 | 41.9 | 137.0 | 477 | 0.813 | 203 / 89 |
| Journal of Parasitology | 508 | 13 | 2.6% | 307 | 23.6 | 80.9 | 293 | 1.405 | -- |
| New England | 1580 | 760 | 48.1% | 12833 | 16.9 | 31.8 | 332 | 53.298 | 2149 / 115180 |
| Journal of Medicine | | | | | | | | | |
| Human & Experimental Toxicology | 406 | 18 | 4.4% | 248 | 13.8 | 52.2 | 223 | 1.772 | -- |
| Nature | 2577 | 1083 | 42.0% | 13430 | 12.4 | 25.2 | 274 | 36.280 | 1035 / 31956 |
| Sexual Plant Reproduction | 77 | 1 | 1.3% | 11 | 11.0 | -- | 11 | 1.869 | -- |
| Quarterly Journal of Experimental Psychology | 375 | 27 | 7.2% | 296 | 11.0 | 22.8 | 87 | 1.964 | -- |
| Science | 3140 | 1081 | 34.4% | 11546 | 10.7 | 23.9 | 558 | 31.201 | 8949 / 109287 |
| Health Physics | 408 | 27 | 6.6% | 269 | 10.0 | 21.1 | 92 | 1.680 | -- |
| Journal of Physical Chemistry A | 3708 | 89 | 2.4% | 853 | 9.6 | 61.1 | 549 | 2.946 | -- |
| Worldviews on Evidence-based Nursing | 63 | 8 | 12.7% | 76 | 9.5 | 18.1 | 54 | 1.239 | -- |
| Learning & Behavior | 76 | 5 | 6.6% | 45 | 9.0 | 13.5 | 33 | 2.000 | -- |
| Rhinology | 199 | 1 | 0.5% | 9 | 9.0 | . | 9 | 1.321 | -- |
| Lancet | 1824 | 686 | 37.6% | 5904 | 8.6 | 21.4 | 419 | 38.278 | 1859 / 68916 |
| American Journal of Clinical Nutrition | 944 | 408 | 43.2% | 3320 | 8.1 | 16.8 | 149 | 6.669 | -- |
| Folia Phoniatrica et Logopaedica | 78 | 1 | 1.3% | 8 | 8.0 | -- | 8 | 1.115 | -- |
| Psychological Science | 557 | 239 | 42.9% | 1849 | 7.7 | 18.3 | 197 | 4.431 | -- |
| Behavior Research Methods | 216 | 11 | 5.1% | 85 | 7.7 | 11.6 | 39 | 2.116 | -- |
| JAMA Internal Medicine *(formerly: Archives of Internal Medicine)* | 643 | 205 | 31.9% | 1568 | 7.7 | 19.5 | 222 | 11.462 | 974 / 5504 |
| JAMA | 1137 | 439 | 38.6% | 3293 | 7.5 | 13.7 | 141 | 30.026 | 4474 / 36339 |
| Nature Biotechnology | 321 | 118 | 36.8% | 871 | 7.4 | 10.1 | 54 | 23.268 | 1185 / 19751 |
| Journal of Sexual Medicine | 921 | 150 | 16.3% | 1089 | 7.3 | 34.1 | 392 | 3.552 | -- |
| Personality and Social Psychology Review | 51 | 25 | 49.0% | 170 | 6.8 | 9.1 | 46 | 6.071 | -- |
| Acta Neurochirurgica | 665 | 24 | 3.6% | 157 | 6.5 | 26.5 | 131 | 1.520 | -- |
| Ophthalmic and Physiological Optics | 202 | 8 | 4.0% | 52 | 6.5 | 12.9 | 38 | 1.583 | -- |

As can be seen in Figure 1 and Table 3, journals that have both high Twitter coverage and high Twitter citation rates are either general science or medical journals, or are related to psychology, nutrition, or sexuality. The high Twitter citation rate of journals with low coverage (*Cutis*, *Journal of Parasitology*, *Human & Experimental Psychology)* is invariably caused by a single article that has been highly tweeted, which may make up as much as 95.4% of the journal's total number of tweets. Interestingly, only 8 of the 25 journals with the highest Twitter citation rates have official Twitter accounts, while 9 of the 13 journals with the highest Twitter coverage ($P\%_{tweeted}$ > 50%) were officially represented on Twitter. However, these Twitter profiles differ in terms of their numbers of tweets and followers.

While the *British Dental Journal* has more than 3,000 followers and 6,000 tweets, *Cardiology in Review* is only followed by 8 users and has tweeted 527 times. *Current Opinion in Cardiology* tweets less (445 tweets) but has a larger direct audience (171 followers).

**Table 4.** Spearman correlations between mean Twitter citation rate for articles tweeted at least once ($T/P_{tweeted}$) and Twitter coverage ($P\%_{tweeted}$) per journal and 2011 Impact Factor (*IF*), Eigenfactor (*EF*) and Article Influence (*AI*) score and Immediacy Index (*II*) for PubMed/WoS articles published in 2011.

| Spearman's ρ | $T/P_{tweeted}$ | $P\%_{tweeted}$ | IF | EF | AI | II |
|---|---|---|---|---|---|---|
| $T/P_{tweeted}$ | 1 | .510** | .242** | .238** | .279** | .247** |
| | N=3725 | N=3725 | N=3628 | N=3628 | N=3628 | N=3628 |
| $P\%_{tweeted}$ | .510** | 1 | .305** | .223** | .312** | .282** |
| | N=3725 | N=3812 | N=3712 | N=3712 | N=3712 | N=3712 |

** Correlation is significant at the 0.01 level (2-tailed).

In order to determine whether a journal's popularity on Twitter is related to its scientific prestige as expressed by citations, Twitter coverage and Twitter citation rates were correlated with the journals' 2011 Impact Factors, Immediacy Indexes, and Eigenfactor and Article Influence Scores (Table 4). All correlations were significant and positive; however, no correlation between the Twitter indicators and journal indicators exceeded .312. In general, correlations were higher with Twitter coverage than with Twitter citation rates.

***Disciplines and specialties.*** As can be seen in Table 5, Twitter coverage at the discipline level is highest in *Professional Fields*, where 17.0% of PubMed documents were mentioned on Twitter at least once, followed by *Psychology* (14.9%) and *Health* (12.8%). When the data set is limited to only those articles that have been tweeted at least once, the papers from *Biomedical Research* have the highest Twitter citation rate ($T/P_{tweeted}$=3.3). Of the 284,764 research articles and reviews assigned to this discipline, 27,878 were mentioned on Twitter a total of 90,633 times. Twitter coverage is lowest in *Physics* papers covered by PubMed (1.8%), and *Mathematics* papers related to biomedical research receive the lowest average number of tweets per tweeted document ($T/P_{tweeted}$ = 1.5).

**Table 5.** Twitter coverage and citations per article by discipline and specialty. Number of papers in PubMed ($P_{PubMed}$), tweeted documents ($P_{tweeted}$), Twitter coverage ($P\%_{tweeted}$), number of tweets (*T*) and mean Twitter citation rate ($T/P_{tweeted}$) per discipline (highlighted according to coloring in Figure 2). Superordinate specialties are listed if more than half of their WoS papers are covered in PubMed. Data for specialties include Twitter citation rate of documents published in 2011 ($T/P^{2011}$), citation rate of tweeted documents published in 2011 ($C/P^{2011}$) and Spearman correlation between the number of tweets and citations per document published in 2011. ID refers to the labels used in Figure 2.

| Disciplines & specialties | $P_{PubMed}$ | $P_{tweeted}$ | $P\%_{tweeted}$ | T | $T/P_{tweeted}$ | $T/P^{2011}$ | $C/P^{2011}$ | Spearman | ID |
|---|---|---|---|---|---|---|---|---|---|
| **Biology** | **61587** | **4357** | **7.1%** | **9634** | **2.2** | | | | |
| Miscellaneous Biology | 5400 | 851 | 15.8% | 2437 | 2.9 | 2.9 | 4.4 | .111* | 59 |
| **Biomedical Research** | **284764** | **27878** | **9.8%** | **90633** | **3.3** | | | | |
| Anatomy & Morphology | 2389 | 138 | 5.8% | 204 | 1.5 | 1.3 | 2.7 | -.100 | 32 |
| Biochemistry & Molecular Biology | 94085 | 7434 | 7.9% | 15194 | 2.0 | 2.3 | 8.5 | .255** | 9 |
| Biomedical Engineering | 16430 | 640 | 3.9% | 895 | 1.4 | 1.4 | 4.0 | .000 | 38 |
| Biophysics | 4648 | 169 | 3.6% | 239 | 1.4 | 1.3 | 4.6 | .000 | 40 |
| Cellular Biology Cytology & Histology | 21184 | 1724 | 8.1% | 3177 | 1.8 | 1.9 | 7.6 | .257** | 29 |
| Embryology | 3211 | 417 | 13.0% | 692 | 1.7 | 1.6 | 5.1 | .205** | 2 |
| Genetics & Heredity | 28164 | 3186 | 11.3% | 7091 | 2.2 | 2.5 | 10.3 | .290** | 46 |
| Microbiology | 29530 | 2009 | 6.8% | 3704 | 1.8 | 1.9 | 5.2 | .146** | 53 |
| Microscopy | 1515 | 47 | 3.1% | 86 | 1.8 | 2.4 | 3.8 | .100 | 19 |
| Miscellaneous Biomedical Research | 8998 | 699 | 7.8% | 1532 | 2.2 | 2.0 | 3.9 | .100 | 5 |
| Nutrition & Dietetic | 12400 | 2534 | 20.4% | 9998 | 3.9 | 4.2 | 3.4 | .179** | 24 |
| Parasitology | 4855 | 182 | 3.7% | 534 | 2.9 | 1.4 | 2.8 | .000 | 54 |
| Physiology | 14677 | 1449 | 9.9% | 4219 | 2.9 | 3.5 | 4.6 | .257** | 1 |
| Virology | 10704 | 896 | 8.4% | 1482 | 1.7 | 1.7 | 3.7 | .100 | 27 |
| **Chemistry** | **121770** | **6640** | **5.5%** | **10933** | **1.6** | | | | |
| Analytical Chemistry | 23256 | 1816 | 7.8% | 2349 | 1.3 | 1.3 | 3.6 | .098** | 55 |
| **Clinical Medicine** | **774961** | **77991** | **10.1%** | **184002** | **2.4** | | | | |
| Addictive Diseases | 5678 | 870 | 15.3% | 2191 | 2.5 | 2.3 | 2.5 | .000 | 28 |
| Allergy | 2608 | 432 | 16.6% | 1381 | 3.2 | 2.8 | 6.3 | .000 | 41 |
| Anesthesiology | 7426 | 1604 | 21.6% | 2626 | 1.6 | 1.5 | 3.0 | .206** | 21 |
| Arthritis & Rheumatology | 9499 | 991 | 10.4% | 1889 | 1.9 | 1.8 | 4.0 | .108* | 39 |
| Cancer | 56138 | 6147 | 10.9% | 12990 | 2.1 | 2.2 | 7.5 | .208** | 18 |
| Cardiovascular System | 42766 | 4004 | 9.4% | 6888 | 1.7 | 1.7 | 6.1 | .177** | 37 |
| Dentistry | 17502 | 979 | 5.6% | 1672 | 1.7 | 1.8 | 1.5 | .000 | 14 |
| Dermatology & Venerial Disease | 15174 | 1270 | 8.4% | 2969 | 2.3 | 1.9 | 2.3 | .166** | 20 |
| Endocrinology | 24680 | 3075 | 12.5% | 7659 | 2.5 | 2.4 | 4.7 | .153** | 17 |
| Environmental & Occupational Health | 10948 | 1455 | 13.3% | 4026 | 2.8 | 2.6 | 3.4 | .187** | 47 |
| Fertility | 6119 | 493 | 8.1% | 1118 | 2.3 | 2.1 | 3.8 | .100 | 12 |
| Gastroenterology | 22911 | 1793 | 7.8% | 3425 | 1.9 | 1.7 | 5.0 | .270** | 26 |
| General & Internal Medicine | 62939 | 8242 | 13.1% | 37479 | 4.5 | 5.8 | 8.7 | .327** | 51 |
| Geriatrics | 3606 | 477 | 13.2% | 1102 | 2.3 | 2.3 | 3.6 | .100 | 6 |
| Hematology | 17354 | 1360 | 7.8% | 2326 | 1.7 | 1.7 | 5.6 | .199** | 4 |
| Immunology | 45162 | 4626 | 10.2% | 9311 | 2.0 | 1.9 | 5.8 | .176** | 11 |
| Miscellaneous Clinical Medicine | 11942 | 2351 | 19.7% | 8411 | 3.6 | 3.6 | 2.3 | .105** | 52 |
| Nephrology | 6615 | 500 | 7.6% | 790 | 1.6 | 1.5 | 4.7 | .100 | 13 |
| Neurology & Neurosurgery | 87449 | 9396 | 10.7% | 22646 | 2.4 | 2.6 | 6.0 | .154** | 23 |
| Obstetrics & Gynecology | 15523 | 1378 | 8.9% | 2582 | 1.9 | 1.8 | 2.6 | .152** | 42 |
| Ophthalmology | 15761 | 1057 | 6.7% | 1633 | 1.5 | 1.4 | 2.8 | .000 | 22 |
| Orthopedics | 16691 | 1618 | 9.7% | 3434 | 2.1 | 2.2 | 2.7 | .177** | 36 |
| Otorhinolaryngology | 11838 | 882 | 7.5% | 1379 | 1.6 | 1.6 | 1.3 | .000 | 8 |
| Pathology | 13110 | 1061 | 8.1% | 1800 | 1.7 | 1.4 | 3.1 | .140** | 10 |
| Pediatrics | 18334 | 2597 | 14.2% | 7312 | 2.8 | 2.9 | 2.9 | .235** | 30 |
| Pharmacology | 69302 | 5232 | 7.5% | 9151 | 1.7 | 1.8 | 4.2 | .110** | 31 |
| Pharmacy | 6726 | 302 | 4.5% | 415 | 1.4 | 1.5 | 1.5 | -.100 | 7 |
| Psychiatry | 20363 | 3453 | 17.0% | 8229 | 2.4 | 2.3 | 4.2 | .157** | 33 |
| Radiology & Nuclear Medicine | 31288 | 2168 | 6.9% | 3235 | 1.5 | 1.4 | 2.4 | .000 | 16 |
| Respiratory System | 9821 | 1292 | 13.2% | 2851 | 2.2 | 2.0 | 5.2 | .093* | 35 |
| Surgery | 48247 | 4038 | 8.4% | 5886 | 1.5 | 1.5 | 2.2 | .185** | 25 |
| Tropical Medicine | 4380 | 305 | 7.0% | 485 | 1.6 | 1.5 | 2.1 | .000 | 61 |
| Urology | 17678 | 1177 | 6.7% | 2799 | 2.4 | 3.0 | 2.9 | .204** | 43 |
| **Earth and Space** | **26925** | **1070** | **4.0%** | **2885** | **2.7** | | | | |
| **Engineering and Technology** | **27567** | **1517** | **5.5%** | **2916** | **1.9** | | | | |
| **Health** | **58580** | **7483** | **12.8%** | **17306** | **2.3** | | | | |
| Geriatrics & Gerontology | 3317 | 344 | 10.4% | 654 | 1.9 | 1.8 | 2.4 | .218** | 56 |
| Health Policy & Services | 11429 | 1774 | 15.5% | 4636 | 2.6 | 2.6 | 2.3 | .116** | 44 |
| Nursing | 13348 | 1579 | 11.8% | 2987 | 1.9 | 1.8 | 0.8 | .125** | 34 |
| Public Health | 18692 | 2289 | 12.2% | 5546 | 2.4 | 2.4 | 2.2 | .078* | 49 |
| Rehabilitation | 7415 | 847 | 11.4% | 1836 | 2.2 | 2.0 | 1.9 | .000 | 48 |
| Social Sciences, Biomedical | 2612 | 486 | 18.6% | 1329 | 2.7 | 2.7 | 2.1 | .100 | 3 |
| Social Studies of Medicine | 190 | 10 | 5.3% | 10 | 1.0 | 1.0 | 0.2 | .000 | 15 |
| Speech-Language Pathology and Audiology | 1577 | 154 | 9.8% | 308 | 2.0 | 1.8 | 1.7 | -.200 | 50 |
| **Humanities** | **691** | **45** | **6.5%** | **121** | **2.7** | | | | |
| **Mathematics** | **2459** | **134** | **5.4%** | **197** | **1.5** | | | | |
| **Physics** | **19353** | **340** | **1.8%** | **539** | **1.6** | | | | |
| **Professional Fields** | **5586** | **950** | **17.0%** | **2510** | **2.6** | | | | |
| **Psychology** | **35873** | **5350** | **14.9%** | **16240** | **3.0** | | | | |
| Behavioral Science & Complementary Psychology | 4802 | 521 | 10.8% | 1385 | 2.7 | 2.7 | 2.6 | .000 | 57 |
| Clinical Psychology | 5757 | 994 | 17.3% | 2161 | 2.2 | 2.1 | 2.4 | .100 | 58 |
| Developmental & Child Psychology | 5502 | 937 | 17.0% | 2365 | 2.5 | 2.5 | 2.7 | .120* | 60 |
| Experimental Psychology | 7502 | 1106 | 14.7% | 4795 | 4.3 | 4.5 | 3.5 | .100 | 45 |
| **Social Sciences** | **8922** | **812** | **9.1%** | **2192** | **2.7** | | | | |
| **Total** | **1431399** | **134929** | **9.4%** | **340751** | **2.5** | | | | |

* Correlation is significant at the 0.05 level (2-tailed).
** Correlation is significant at the 0.01 level (2-tailed).

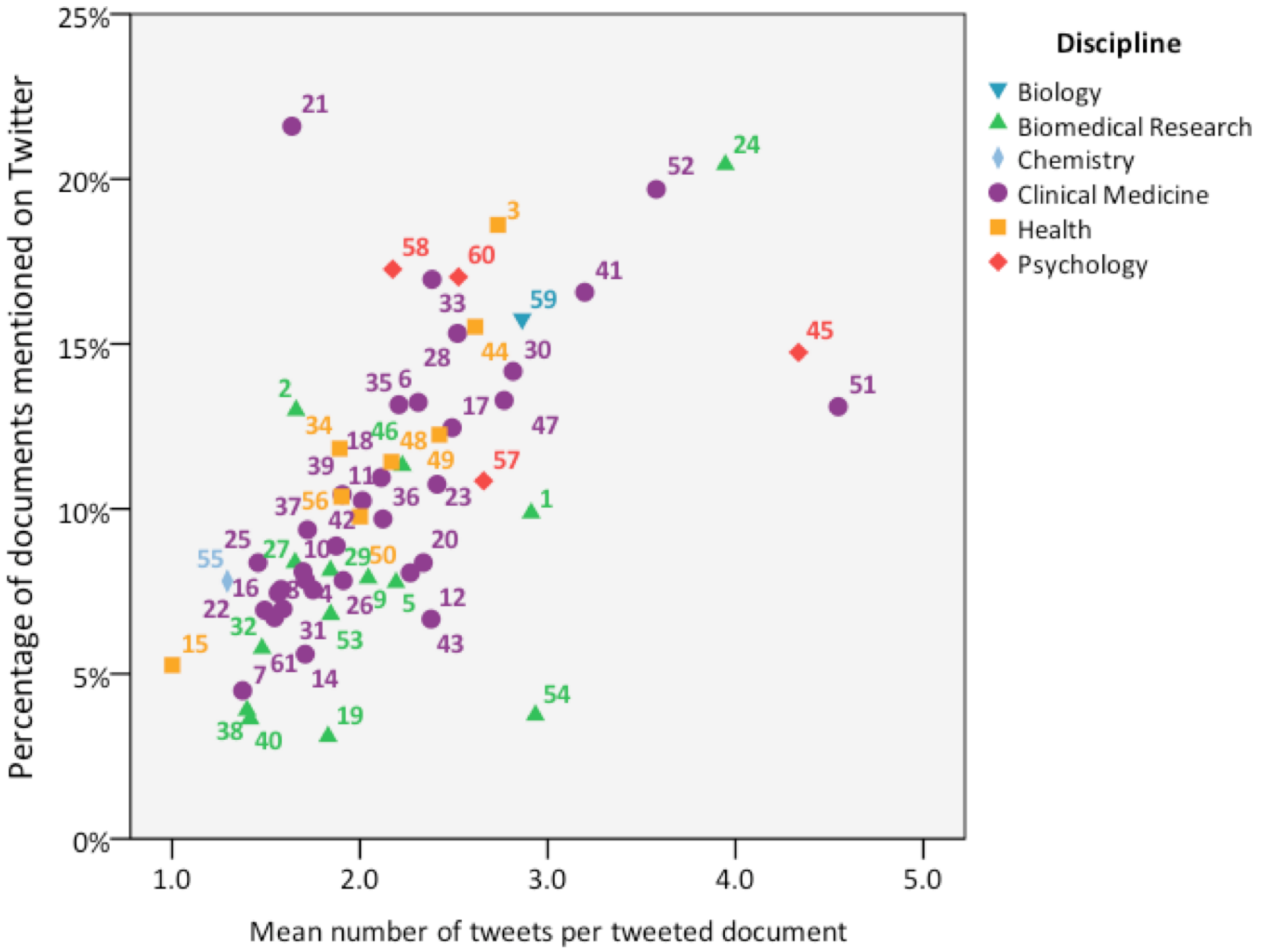

**Figure 2.** Percentage of documents per specialty mentioned on Twitter (Twitter coverage) and mean number of tweets per tweeted document (Twitter citation rate) for the 61 specialties with PubMed coverage of at least 50%. Coloring indicates superordinate discipline and labels show specialty IDs as listed in Table 5.

On the level of specialties, *General & Internal Medicine* (ID in 2: 51) has the highest Twitter citation rate and a coverage rate of 13.1%. Tweeted articles from this specialty on Twitter are (re)tweeted 4.5 times on average (Figure 2, Table 5). The specialty is popular among Twitter users in terms of diversity (number of different papers tweeted) and popularity (number of tweets per tweeted document) compared to the average Twitter coverage ($P\%_{tweeted}$=9.4%) and Twitter citation rate ($T/P_{tweeted}$=2.5) of all PubMed papers. The same is true for *Experimental Psychology* ($T/P_{tweeted}$=4.3; $P\%_{tweeted}$=14.7%; ID: 45), *Nutrition & Dietics* (3.9; 20.4%; ID: 24), *Miscellaneous Clinical Medicine* (3.6; 19.7%; ID: 52) and *Allergy* (3.2; 16.6%; ID: 41). With more than a fifth of its papers mentioned at least once on Twitter, *Anesthesiology* (1.6; 21.6%; ID: 21) has the highest coverage but is tweeted below average.

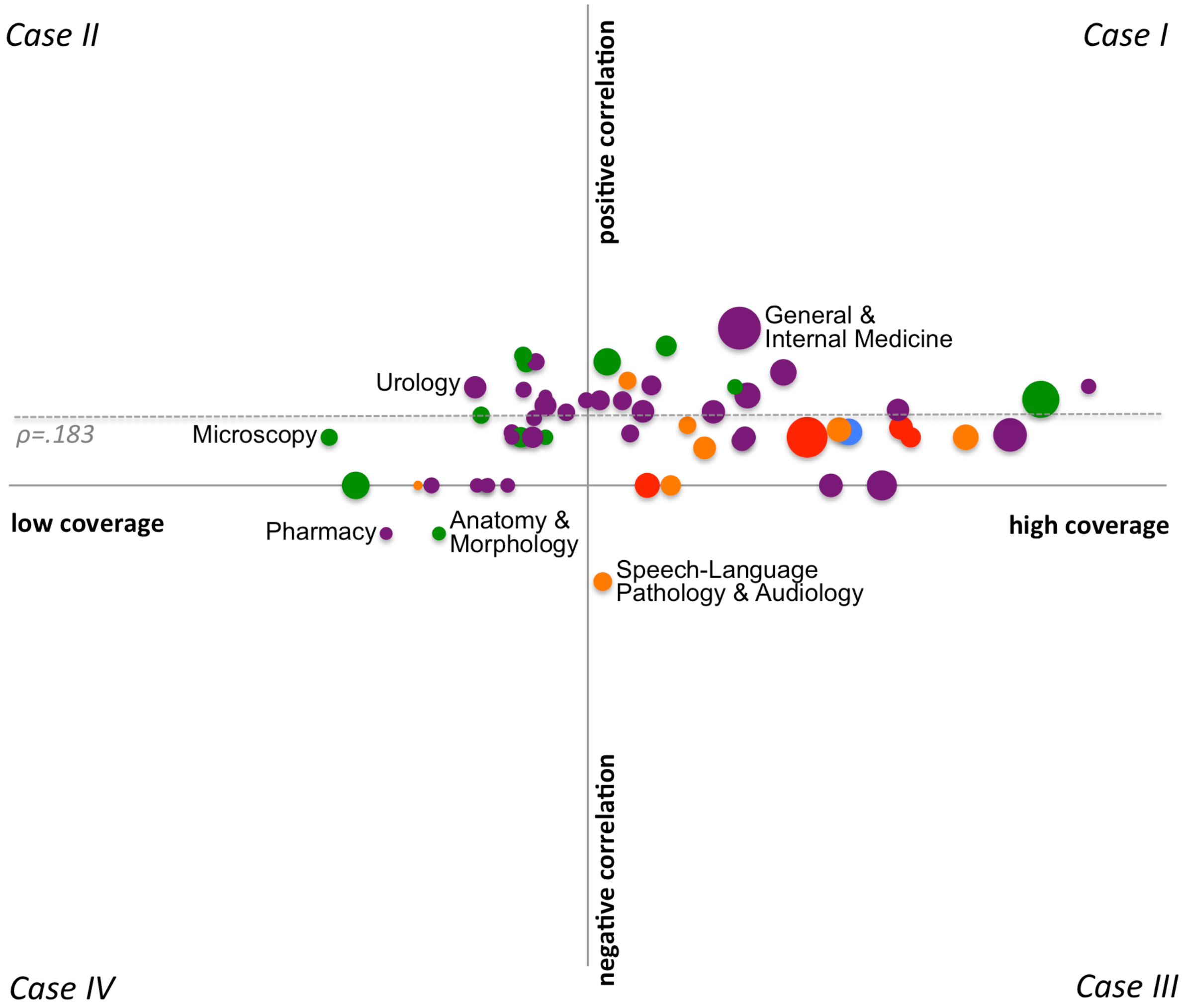

**Figure 3.** Specialties assigned to the four cases reflecting positive and negative correlations values between citations and tweets per document and high and low coverage of documents on Twitter. Coloring indicates superordinate discipline as listed in Table 5 and size of data points the Twitter citation rate. Coverage values are normalized by the average for all PubMed documents (9.4%), so that data points placed in the first and second quadrant represent specialties with positive correlations and coverage values above (*Case I*) and below average (*Case II*), respectively. The third and fourth quadrants contain those specialties where correlations were negative and coverage above (*Case III*) and below average (*Case IV*). Dashed line represents Spearman correlation between citations and tweets of all 2011 papers (ρ=.183**).

To compare the relationship between tweeting and citation behavior, the number of tweets and citations were analyzed on the document level for each of the 61 specialties. As described in the Methods section, the correlations are based on 2011 publications to reduce citation delay and Twitter uptake biases as far as possible. As shown in Table 5, correlations were positive for most (i.e., 47) of the 61 specialties but very low in general[iv] with *General & Internal Medicine* having the highest Spearman values of .327. While 13 specialties showed no correlation between tweets and citations, the Spearman value was negative in three specialties, i.e. *Speech-Language Pathology & Audiology* (ρ=-.200), *Anatomy & Morphology* (ρ=-.100) and *Pharmacy* (ρ=-.100). Correlations were significant for only 26 specialties, all of which were positive. Even where correlations are the highest, there are huge differences in the ranking positions of individual papers. For example, the three most frequently tweeted papers were ranked 161$^{st}$, 1996$^{th}$ and 1007$^{th}$ in terms of citations within the *General & Internal Medicine* specialty in our dataset and the three most frequently cited papers in the specialty were ranked 610$^{th}$, 18$^{th}$ and 228$^{th}$ by number of tweets. The second most frequently cited (443 citations) and 18$^{th}$ most frequently tweeted document (123 tweets) discusses antiretroviral therapy to limit the transmission of HIV and is consequentially highly relevant from both a social and medical research perspective, and the *General & Internal Medicine* paper that ranked highest in terms of tweets and citations concerns a topic that has major implications for physicians treating stroke risk patients and a large community of patients themselves (see Mohammadi & Thelwall, 2013, in press).

Figure 3 depicts coverage and correlations between number of citations and number of tweets. Note that coverage values are normalized by the average value for the entire data set (9.4%) so that data points placed in the first quadrant represent specialties with positive correlations and coverage above average (*Case I*), suggesting that tweeting and citation behavior tend to overlap more and that documents are represented more broadly than average, e.g. *General & Internal Medicine* (ρ= .327**, $P\%_{tweeted}$=13.1%). Specialties in the second quadrant reflect positive correlations and coverage below average (*Case II*), such as *Urology* (ρ=.204**, $P\%_{tweeted}$=6.7%) and *Microscopy* (ρ=.100, $P\%_{tweeted}$=3.1%), which are less popular on Twitter. While the third quadrant contains those specialties with negative correlations and coverage above average (*Case III*), indicating that the specialties are popular on Twitter but citation and tweeting behaviour differs (*Speech-Language Pathology & Audiology* (ρ=-.200, $P\%_{tweeted}$=9.8%), the fourth quadrant contains those with negative correlations and coverage values below average (*Case IV*), i.e. *Anatomy & Morphology* (ρ=-.100, $P\%_{tweeted}$=5.8%) and *Pharmacy* (ρ=-.100, $P\%_{tweeted}$=4.5%). Size of data points represents the Twitter citation rate per discipline.

## Discussion and Framework

***Discussion***. The goal of this paper was to examine the degree to which biomedical papers appeared on Twitter, the degree to which this varied by journal and domain, and the relationship between tweets and citations. We briefly discuss our findings in relationship to these three areas.

Less than 10% of the more than 1.4 million articles found in both WoS and PubMed were tweeted. However, Twitter coverage has increased dramatically over time; with more than 20% of articles published in 2012 receiving at least on tweet. These rates of coverage are much lower than those found for other sources of altmetric data, such as the readership data generated from Mendeley (e.g., Bar-Ilan, Haustein, Peters, Priem, Shema, & Terliesner, 2012; Haustein, Peters, Bar-Ilan, Priem, Shema, & Terliesner, 2013; Bar-Ilan, 2012a; 2012b; Li et al., 2012). The majority of journals had less than 20% of their content tweeted. Those with high Twitter coverage tended to be those with designated Twitter handles for the journal or the associated publisher/association. However, the maintenance of an official Twitter account did not necessarily translate to an increased Twitter citation rate for articles within these journals. On average, articles which were tweeted were tweeted two and half times, though most only received a single tweet. At 0.2 tweets per article, the overall Twitter citation rate was significantly lower when including untweeted articles. Wide variation was also found by discipline and specialty in both Twitter coverage and Twitter citation rate. Results strongly show that, as with publication and citation behavior, tweeting behavior varies across disciplines and specialties, and these differences need to be accounted for when comparing the social media impacts of scholarly articles from different fields.

Correlations between Twitter coverage and Twitter citation rates with traditional bibliometric indicators for journals were positive and significant, with rates between .223 and .312. Comparing formal citations and Twitter citations for all papers published in 2011, we found a low but positive correlation of .183, which suggests that, although both indicators are somewhat related, they mostly measure a different type of impact. Moreover, these correlations are lower than the coefficients identified relating other novel metrics (such as readership and mentorship) to citations (Schlögl, Gorraiz, Gumpenberger, Jack, & Kraker, 2013; Bar-Ilan et al., 2013; Li et al., 2012; Sugimoto, Russell, Meho, & Marchionini, 2008) and are also lower than the demonstrated correlations between tweets and other metrics, such as downloads and Google Scholar citations (Shuai et al., 2012).

Given the low correlations found here on a very large dataset of both tweets and citations and limiting the biases of Twitter uptake on the one hand and citation delays on the other, we argue that Twitter citations do not to reflect traditional research impact. This may be due to several factors, including the low Twitter uptake among scientists and the fact that the viability of Twitter as a tool for scholarly communication is still mostly unknown. Reasons for low Twitter usage among scholars will need to be determined in future qualitative usage studies.

That being said, our exploratory analysis of top tweeted articles suggests that they might actually have been highly tweeted because of their curious or humorous content, implying that these tweets are mostly made by the "general public" rather than the scientific community. In other words, their high number of tweets did not seem to be due to their intellectual contribution or scientific quality. Other articles were highly tweeted because of their timeliness or their health-related content, which is the closest we've been to assessing what one could consider as the impact of health research on society. All in all, these findings suggests that there is a lot of heterogeneity in the tweeting of biomedical papers, and that a lot of work still needs to be done in order to assess the various contexts in which scientific papers are tweeted. More specifically, it is of crucial importance to obtain robust data on the relative importance of each of these contexts to see, among other things, whether tweets are indeed a measure of the social impact of research, before these metrics are added to the scientometric toolbox.

***Framework***. To facilitate interpretation of results, we present a framework for understanding the relationships among new and established indicators. For the present, we focus on Twitter as a case study and use our data to inform the framework. However, this could be easily expanded to encompass other new indicators. Although we believe that various altmetrics differ and that social media based indicators comprise many different facets that should not be blended but analyzed separately. For example, as a microblogging platform Twitter can be considered a tool for dissemination and discussion, Mendeley most likely reflects readership within the academic community, while F1000 provides post-publication peer review. Just like citations, downloads and standard peer-reviews, these aspects should be considered separately and not be aggregated into one single number or indicator as impact is a multifaceted notion (Haustein, 2012).

Figure 4 provides a schematic representation of the framework of tweeting behavior of scholarly documents, which is divided into four cases according to their 1) Twitter coverage and 2) correlation between number of Tweets and citations.

- *Case I: A set of documents (e.g., journals, specialties, disciplines) has high Twitter coverage, papers with many Twitter citations have many WoS citations, and papers with few Twitter citations have few citations (high positive correlation)*
  The coverage suggests either that there is a large user group on Twitter mentioning scholarly documents or there is systematic tweeting by a few individuals or automated tweeting. A high correlation suggests that Twitter users are interested in the same papers as the scientific community since documents that have a high citation impact are also popular on Twitter and those that are not frequently cited are also tweeted about less often. This scenario is consistent with the following.
  Many members of the scientific community and/or the general public are active on Twitter and tweet a large proportion of published scientific articles AND
    A) The scientific community discusses many different scientific articles on Twitter, so that Twitter serves as an alternative way to find, distribute and discuss results within the scientific community. Scientific tweeting patterns follow traditional citing patterns.
    AND/OR
    B) The general public discusses scholarly findings on Twitter and is interested in the same topics as scientists (e.g., general interest topics in health related issues) and tweets in the same way that scientists cite their publications.
  As demonstrated in our study, high coverage does not necessarily imply that there is great interest in the articles. Rather, it could imply that tweeting has been incorporated into the professional activities of the journal or publisher. Regardless, high coverage indicates high dissemination on the platform. A direct correlation between the new metric (tweets, in this case) and an established metric (e.g., citations) suggests that these metrics are reinforcing and consensual. In the case that the metrics represent separate audiences, we might infer broader impact. If those generating the metrics are the same population, we might infer redundancy (e.g., if the same scientists are tweeting and citing, no added measure of impact can be derived).
- *Case II: A set of documents has low Twitter coverage, papers with many Twitter citations have many WoS citations, papers with many Twitter citations have many WoS citations, and papers with few Twitter citations have few WoS citations (high positive correlation)*
  The coverage shows that the user group on Twitter is only interested in a few scholarly documents from this set. The correlation suggests that Twitter users are interested in the same topics and papers as the scientific community since documents that have a high citation impact are also popular on Twitter. Scientific papers that have low impact on the scholarly community are not popular on Twitter. The scenarios causing this conclusion are consistent with the following.
  Many members of the scientific community and/or the general public are active on Twitter but tweet only a small proportion of published scientific articles AND
    A) The scientific community actively discusses a few scientific articles on Twitter.
    AND/OR

B) The general public is interested in topics (e.g., medical health related issues) and tweets about scientific papers.

- *Case III: A set of documents has high Twitter coverage, papers with few Twitter citations have many WoS citations, and papers mentioned frequently on Twitter have few WoS citations (high negative correlation)*

  The coverage shows that there is a large user group on Twitter mentioning scholarly documents but the correlation reflects that Twitter users are not interested in the same articles as the scientific community, since papers that have a high scientific impact are not popular on Twitter. The scenarios which cause this conclusion are consistent with the following.

  A large share of (scientific and/or general public) community is active on Twitter and distributes a large share of documents AND

  A) The scientific community tweets different content than it cites in scholarly publications.

  AND/OR

  B) The general public is not interested in the same topics as scientists cite but is interested in articles that have low scientific impact (e.g., curious papers with funny titles).

  If those who are tweeting are also citing, then this implies that the metrics measure different types of use or interest in these articles. For example, medical researchers may tweet and rate highly articles that help clinical practice even if they do not then cite them (e.g., for related results see Mohammadi & Thelwall, 2013, in press). An alternative explanation is that the user groups are different—that is, that those who tweet do not also write and, thereby, cite in this area. One potential scenario in this group are curious papers with particularly provocative or interesting titles.

- *Case IV: A set of documents has low Twitter coverage, papers with few Twitter citations have many WoS citations, and papers mentioned frequently on Twitter have few WoS citations (high negative correlations)*

  The coverage shows that the user group on Twitter is only interested in a few scholarly documents and the correlation reflects that Twitter users are not interested in the same topics as the scientific community, since papers that have a high scientific impact are not popular on Twitter. The scenarios which cause this conclusion are consistent with the following.

  Only a small share of documents is distributed by the (scientific and/or general public) community (either a small or large share of the respective community) AND

  A) The scientific community tweets different content than it cites in scholarly publications.

  AND/OR

  B) The general public is not interested in the same topics as scientists cite but is interested in articles that have low scientific impact (e.g., curious papers with funny titles).

  This differentiates from Case III, in that the correlations are generated by very few articles—that is, an article that is either very highly cited or highly tweeted. A negative correlation but low coverage might imply that while there is only a small community on Twitter which is not capable of distributing a large amount of different articles, those papers that *are* distributed are relevant to both scientists and a general public.

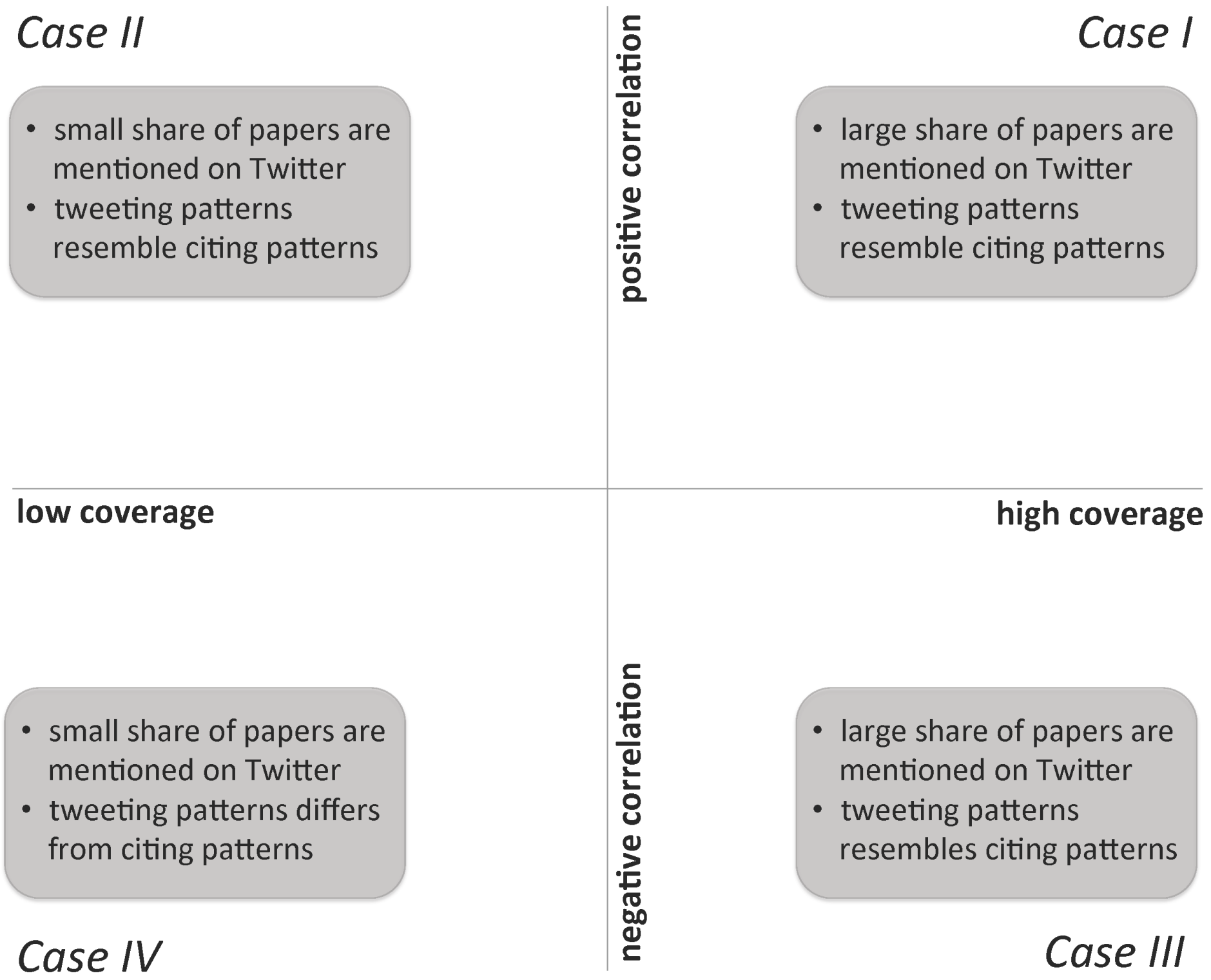

**Figure 4.** Framework of *Cases I* to *IV* representing high and low Twitter coverage and correlation between the number of tweets and citations for a set of documents.

As demonstrated by these cases, there are many diverse actors, motivations, processes, and outcomes embedded in interpretations of altmetric data. Qualitative and more detailed analyses of Twitter users are needed to determine whether scenarios A and/or B apply or whether it is a completely different group of actors responsible for the distribution of scholarly literature on Twitter. Further research should thus include qualitative analyses of Twitter content as well as user surveys aimed at determining Twitter's role in the scholarly communication and reward systems and shed light on researchers' motivations for (not) using Twitter. In addition, analyses such as ours should be performed for non-medical disciplines so that more can be known about discipline-specific Twitter uptake and its appropriateness as a tool in research evaluation.

## Conclusion

This large-scale analysis covering the entire spectrum of medical disciplines provides substantial data for the evaluation of Twitter metrics and supports the understanding of scholarly Twitter use in this research area. We introduced a framework to classify four distinct relationships between tweeting and citing, using this as a theoretical underpinning to facilitate the evaluation of tweeting behavior in various specialties.

With less than 10% of PubMed documents mentioned, Twitter shows a much lower coverage of scholarly document than other social media platforms such as Mendeley and CiteULike, which is most likely due to the scholarly focus of the latter two. Nevertheless, we were able to demonstrate that there are some journals and specialties in biomedical science that are of greater interest to the Twitter community than others. Low correlation between the number of citations and tweets per document indicate that tweets and citations are far from measuring the same impact and suggest that Twitter-based indicators reflect another kind of impact not comparable to traditional citation indicators. Therefore, they should not be considered as alternatives to citation-based indicators, but rather as complementary.

However, before adding tweets to the scientometric toolbox as a complementary measure of public or societal impact of scholarly products, motivations to tweet need to be distinguished and further evaluated. As shown, the distribution of academic articles on Twitter is in general quite low. This low coverage rate may reflect a belief among the majority of academics that it would not be a good use of their time to tweet about publications, perhaps because interested scholars could find their articles in other ways and do not use Twitter to discuss research. Further research needs to investigate why academics tweet or do not tweet about publications and who the users are that mentioning academic articles on Twitter.

Our exploratory analysis of highly tweeted documents shows that while some papers seem to receive attention on Twitter because of actual health implications or topicality, others seem to be distributed on Twitter due to humorous or curious contents, which suggests that tweets do not necessarily reflect intellectual impact. A large-scale user survey could reveal why scholarly publications are mentioned on Twitter and why some papers are tweeted more frequently than others. Moreover, distribution of scholarly documents on Twitter is influenced by officially curated Twitter handles and particular journal policies. These ambiguities have implications for the use of tweet counts as altmetrics.

**Acknowledgements**
This research was part of the international Digging into Data program (funded by AHRC/ESRC/JISC (UK), SSHRC (Canada), and the National Science Foundation (US; grant #1208804). The authors would like to thank Euan Adie of Altmetric.com for supplying the data and descriptions of it and to Andrew Tsou for work on earlier versions of the manuscript. VL acknowledges funding from the Canada Research Chair program.

---

[i] Altmetric.com and ImpactStory.org collect and sell altmetrics data for scholarly products and offer free statistics for single documents through their websites.

[ii] See Mahrt, Weller, and Peters (2013, in press) for an overview of studies concerning scholars and Twitter.

[iii] http://twitpic.com.

[iv] In order to assess whether low correlations are caused by the citation uptake being slower in some specialties than others, we ran a Pearson correlation between the citation rate and the (Spearman) correlation between tweets and citations. A Pearson value of r=.567 for the 61 specialties shows that a higher tweet/citation correlation is to some extent associated with higher citation rates and might therefore be caused by faster citation uptakes. Specialties that have weaker correlations might not have collected enough citations from their publication in 2011 to the end of 2012, and correlations between tweets and citations might increase with a longer citation window or in disciplines where citation uptake is faster (see Shuai, Pepe, & Bollen, 2012).